\documentclass[aps,prl,twocolumn,superscriptaddress,showpacs,preprintnumbers]{revtex4}
\usepackage{epsfig}
\usepackage{graphicx}
\usepackage{graphics}
\usepackage{dcolumn}
\usepackage{bm}
\usepackage{afterpage}

\newcommand{\etal}{{\it et al.}}

\begin{document}

\preprint{\tighten\vbox{\hbox{\hfil CLNS 07/1989}
                       \hbox{\hfil CLEO 07-01}
}}

\title{Measurement of the Decay Constant
$f_{D_s^+}$ using $D_s^+\to\ell^+\nu$}

\author{M.~Artuso}
\author{S.~Blusk}
\author{J.~Butt}
\author{S.~Khalil}
\author{J.~Li}
\author{N.~Menaa}
\author{R.~Mountain}
\author{S.~Nisar}
\author{K.~Randrianarivony}
\author{R.~Sia}
\author{T.~Skwarnicki}
\author{S.~Stone}
\author{J.~C.~Wang}
\affiliation{Syracuse University, Syracuse, New York 13244}
% ...
\collaboration{CLEO Collaboration} 
\noaffiliation

\date{\today}

\begin{abstract}

We measure the decay constant $f_{D_s^+}$ using the $D_s^+\to
\ell^+\nu$ channel, where the $\ell^+$ designates either a $\mu^+$
or a $\tau^+$, when the $\tau^+\to\pi^+\overline{\nu}$. Using both
measurements we find $f_{D_s^+}=274\pm 13 \pm 7 {~\rm MeV}$.
Combining with our previous determination of $f_{D^+}$, we compute
the ratio $f_{D_s^+}/f_{D^+}=1.23\pm 0.11\pm 0.04$. We compare with
theoretical estimates.
\end{abstract}


\maketitle \tighten

To extract precise information on the size of CKM matrix elements
from $B_d$ and $B_s$ mixing measurements the ratio of ``decay
constants," that are related to the heavy and light quark
wave-function overlap at zero separation, must be well known
\cite{formula-mix}. Recent measurement of $B_s^0$ mixing by CDF
\cite{CDF} has shown the urgent need for precise numbers. Decay
constants have been calculated for both $B$ and $D$ mesons using
several methods, including lattice QCD \cite{Davies}. Here we
present the most precise measurement to date of $f_{D_s^+}$, and
combined with our previous determination of $f_{D^+}$
\cite{our-fDp,DptomunPRD}, we find $f_{D^+_s}/f_{D^+}$.

In the Standard Model (SM) purely leptonic $D_s$ decay proceeds via
annihilation through a virtual $W^+$. The decay rate is given by
\cite{Formula1}
\begin{equation}
\label{eq:formula1}
\small{ \Gamma(D_s^+\to \ell^+\nu) =
{{G_F^2}\over 8\pi}f_{D_s^+}^2m_{\ell}^2M_{D_s^+}
\left(1-{m_{\ell}^2\over M_{D_s^+}^2}\right)^2
\left|V_{cs}\right|^2,} \label{eq:equ_rate}
\end{equation}
where $M_{D_s^+}$ is the $D_s^+$ mass, $m_{\ell}$ is the lepton
mass, $G_F$ is the Fermi constant, and $|V_{cs}|$ is a CKM matrix
element with a value of 0.9738 \cite{PDG}.

In this Letter we report measurements of both
${\cal{B}}(D_s^+\to\mu^+\nu)$ and ${\cal{B}}(D_s^+\to\tau^+\nu)$,
when $\tau^+\to \pi^+\overline{\nu}$
($D_s^+\to\pi^+\overline{\nu}\nu$). More details are given in a
companion paper \cite{PRD}. The ratio
$\Gamma(D_s^+\to\tau^+\nu)/\Gamma(D_s^+\to\mu^+\nu)$ predicted in
the SM via Eq.~\ref{eq:formula1} depends only on well-known
masses, and equals 9.72; any deviation would be a manifestation of
new physics as it would violate lepton universality \cite{Hewett}.
New physics can also affect the expected widths; any undiscovered charged bosons would
interfere with the SM $W^+$ \cite{Akeroyd}.

The CLEO-c detector \cite{CLEODR} is equipped to measure the momenta
of charged particles, identify them using $dE/dx$ and Cherenkov
imaging (RICH) \cite{fakes}, detect photons and determine their
directions and energies. We use 314 pb$^{-1}$ of data produced in
$e^+e^-$ collisions using CESR near 4.170 GeV. Here the
cross-section for our analyzed sample,
$D_s^{*+}D_s^-$+$D_s^{+}D_s^{*-}$, is $\sim$1 nb. Other charm
production totals $\sim$7 nb \cite{poling}, and the underlying
light-quark ``continuum" is $\sim$12 nb. We fully reconstruct one
$D_s^-$ as a ``tag," and examine the properties of the $D_s^+$.
(Charge conjugate decays are used.) Track selection, particle
identification, $\pi^0$, $\eta$, and $K_S^0$ criteria are the same
as those described in Ref. \cite{our-fDp}, except that RICH
identification now requires a minimum momentum of 700 MeV/$c$.

\begin{figure}[hbtp]
\centering
\includegraphics[width=3in]{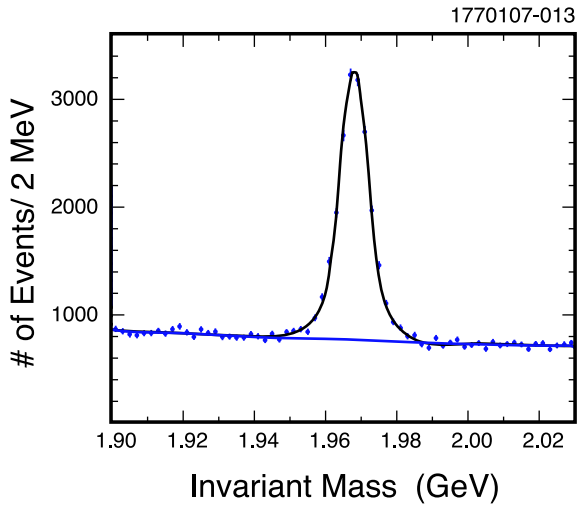}\vskip -2mm
\caption{Invariant mass of $K^+K^-\pi^-$ candidates after requiring
the total energy to be consistent with the beam energy. The curve
shows a fit to a two-Gaussian signal function plus a polynomial
background.
 } \label{Inv-mass}
\end{figure}

Tag modes are listed in Table~\ref{tab:Ntags}. For resonance decays
we select intervals in invariant mass within $\pm$10 MeV of the
known mass for $\eta'\to\pi^+\pi^-\eta$, $\pm$10 MeV for $\phi\to
K^+ K^-$, $\pm$100 MeV for $K^{*0}\to K^-\pi^+$, and $\pm$150 MeV
for $\rho^-\to \pi^-\pi^0$. We require tags to have momentum
consistent with coming from $D_sD_s^*$ production. The distribution
for the $K^+K^-\pi^-$ mode (44\% of all the tags) is shown in
Fig.~\ref{Inv-mass}.

\begin{table}[htb]
\begin{center}
\caption{Tagging modes and numbers of signal and background events,
within cuts, from two-Gaussian fits to the invariant mass plots, and
the number of $\gamma$ tags in each mode, within $\pm2.5\sigma$ from
a fit to the signal Crystal Ball function (see text) and a 5th order
Chebychev background polynomial and the associated
background.\label{tab:Ntags}}
\begin{tabular}{lcccc}
 \hline\hline
   Mode  &   \multicolumn{2}{c}{Invariant Mass}          &  \multicolumn{2}{c}{MM$^{*2}$} \\
       &  Signal          &  Bkgrnd  & Signal & Bkgrnd\\ \hline
$K^+K^-\pi^- $ & 13871$\pm$262 & 10850 & 8053$\pm$ 211 &13538\\
$K_S^0 K^-$ & 3122$\pm$79 & 1609 & 1933$\pm$88&2224\\
$\eta\pi^-$& $1609\pm 112$  &
4666&1024$\pm$97 &3967\\
$\eta'\pi^-$  & 1196$ \pm $46  &409 &792$\pm$69 &1052 \\
$\phi\rho^-$ & 1678$ \pm $74  &1898&1050$\pm$113&3991 \\
$\pi^+\pi^-\pi^-$ & 3654$ \pm $199  & 25208 & 2300$\pm$187& 15723\\
$K^{*-}K^{*0}$ & 2030$
\pm$98& 4878&1298$\pm$130 & 5672\\
$\eta\rho^-$ & 4142$ \pm $281  &20784 & 2195$\pm$225 & 17353\\
\hline
Sum &  $31302\pm 472 $ &70302 & 18645$\pm$426&63520\\
\hline\hline
\end{tabular}
\end{center}
\end{table}

To select tags, we first fit the invariant mass distributions to the
sum of two Gaussians centered at $M_{D_s}$. The r.m.s. resolution
($\sigma$) is defined as $\sigma \equiv
f_1\sigma_1+(1-f_1)\sigma_2$, where $\sigma_1$ and $\sigma_2$ are
the individual widths and $f_1$ is the fractional area of the first
Gaussian. We require the invariant masses to be within $\pm
~2.5\sigma$ ($\pm2\sigma$ for the $\eta\rho^-$ mode) of $M_{D_s}$.
We have a total of 31302$\pm$472 tag candidates. Then we add a
$\gamma$ candidate that satisfies our shower shape requirement.
Regardless of whether or not the $\gamma$ forms a $D_s^*$ with the
tag, for real $D_s^*D_s$ events, the missing mass squared,
MM$^{*2}$, recoiling against the $\gamma$ and the $D_s^-$ tag should
peak at $M^2_{D_s^+}$. We calculate
\[
{\rm MM}^{*2}=\left(E_{\rm CM}-E_{D_s}-E_{\gamma}\right)^2-
\left(\overrightarrow{p}_{\!\rm
CM}-\overrightarrow{p}_{\!D_s}-\overrightarrow{p}_{\!\gamma}\right)^2\!\!,
\]
where $E_{\rm CM}$ ($\overrightarrow{p}_{\!\rm CM}$) is the
center-of-mass energy (momentum), $E_{D_s}$
($\overrightarrow{p}_{\!D_s}$) is the energy (momentum) of the
fully reconstructed $D_s^-$ tag, $E_{\gamma}$
($\overrightarrow{p}_{\!\gamma}$) is the energy (momentum) of the
additional $\gamma$. We use a kinematic fit that constrains the
decay products of the $D_s^-$ to $M_{D_s}$ and conserves overall
momentum and energy. All $\gamma$'s in the event are used, except
for those that are decay products of the $D_s^-$ tag.

\begin{figure}[hbtp]
\centering \vskip -2mm
\includegraphics[width=3in]{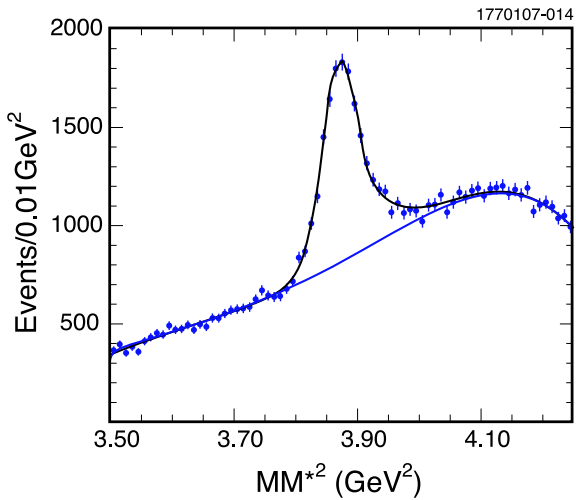}\vskip -2mm
\caption{The MM$^{*2}$ distribution from events with a $\gamma$ in
addition to the $K^+K^-\pi^-$ tag. The curve is a fit to the Crystal
Ball function and a 5th order Chebychev background function.}
\label{MMstar2}
\end{figure}

The MM$^{*2}$ distribution from $K^+K^-\pi^-$ tags is shown in
Fig.~\ref{MMstar2}. We fit all the modes individually to determine
the number of tag events. This procedure is enhanced by having
information on the shape of the signal function. We use fully
reconstructed $D_s^- D_s^{*+}$ events, and examine the signal
shape when one $D_s$ is ignored. The signal is fit to a Crystal
Ball function \cite{taunu}, which determines $\sigma$ and the
shape of the tail. Though $\sigma$ varies somewhat between modes,
the tail parameters don't change, since they depend on beam
radiation and $\gamma$ energy resolution.

Fits of MM$^{*2}$ in each mode when summed show 18645$\pm$426
events within a $\pm 2.5\sigma$ interval (see
Table~\ref{tab:Ntags}). There is a small enhancement of $(4.8 \pm
1.0)$\% in our ability to find tags in $\mu^+\nu$ (or
$\pi^+\overline{\nu}\nu$) events (tag bias) as compared with
generic events. Additional systematic errors are evaluated by
changing the fitting range, using 4th and 6th order Chebychev
background polynomials, and allowing the parameters of the tail of
the fitting function to float, leading to an overall systematic
uncertainty of 5\%.

Candidate $\mu^+\nu$ events are required to have only a single
additional track oppositely charged to the tag with an angle
$>$35.9$^{\circ}$ with respect to the beam line. We also require
that there not be any neutral energy cluster detected of more than
300 MeV, which is especially useful to reject $D_s^+\to \pi^+\pi^0$
and $\eta\pi^+$ decays. Since here we are searching for events in
which there is a single missing $\nu$, the missing mass squared,
MM$^2$, should peak at zero:
\begin{eqnarray}
\label{eq:mm2} {\rm MM}^2=\left(E_{\rm
CM}-E_{D_s}-E_{\gamma}-E_{\mu}\right)^2  \\\nonumber
           -\left(\overrightarrow{p}_{\!\rm CM}-\overrightarrow{p}_{\!D_s}
           -\overrightarrow{p}_{\!\gamma}
           -\overrightarrow{p}_{\!\mu}\right)^2,
\end{eqnarray}
where $E_{\mu}$ ($\overrightarrow{p}_{\!\mu}$) are the energy
(momentum) of the candidate $\mu^+$ track.

We also make use of a set of kinematical constraints and fit each
event to two hypotheses: (1) the $D_s^-$ tag is the daughter of a
$D_s^{*-}$ and (2) the $D_s^{*+}$ decays into $\gamma D_s^+$. The
kinematical constraints, in the center-of-mass frame, are
$\overrightarrow{p}_{\!D_s}+\overrightarrow{p}_{\!D_s^*}=0,~E_{\rm
CM}=E_{D_s}+E_{D_s^*},~E_{D_s^*}={E_{\rm
CM}}/{2}+\left({M_{D_s^*}^2-M_{D_s}^2}\right)/{2E_{\rm
CM}}{\rm~or~} E_{D_s}={E_{\rm
CM}}/{2}-\left({M_{D_s^*}^2-M_{D_s}^2}\right)/{2E_{\rm
CM}},~M_{D_s^*}-M_{D_s}=143.6 {\rm ~MeV}.$ In addition, we
constrain the invariant mass of the $D_s^-$ tag to $M_{D_s}$. This
gives a total of 7 constraints. The missing $\nu$ four-vector
needs to be determined, so we are left with a three-constraint
fit. We perform an iterative fit minimizing $\chi^2$. To eliminate
systematic uncertainties that depend on understanding the absolute
scale of the errors, we do not make a $\chi^2$ cut but simply
choose the $\gamma$ and the decay sequence in each event with the
minimum $\chi^2$.

We consider three separate cases: (i) the track deposits $<$~300 MeV
in the calorimeter, characteristic of a non-interacting pion or a
$\mu^+$; (ii) the track deposits $>$~300 MeV in the calorimeter,
characteristic of an interacting pion; or (iii) the track satisfies
our electron selection criteria. The separation between muons and
pions is not complete. Case (i) contains 99\% of the muons but also
60\% of the pions, while case (ii) includes 1\% of the muons and
40\% of the pions \cite{DptomunPRD}. Case (iii) does not include any
signal but is used for background estimation. For cases (i) and (ii)
we insist that the track not be identified as an electron or a kaon.
Electron candidates have a match between the momentum measured in
the tracking system and the energy deposited in the CsI calorimeter,
and $dE/dx$ and RICH measurements consistent with this hypothesis.

For the $\mu^+\nu$ final state the MM$^2$ distribution is modeled as
the sum of two Gaussians centered at zero. A Monte Carlo (MC)
simulation of the MM$^2$ shows $\sigma$=0.025 GeV$^2$ after the fit.
We check the resolution using the $D_s^+\to \overline{K}^0K^+$ mode.
We search for events with at least one additional track identified
as a kaon using the RICH detector, in addition to a $D_s^-$ tag. The
MM$^2$ resolution is 0.025 GeV$^2$ in agreement with the simulation.

In the $\pi^+\overline{\nu}\nu$ final state, the extra missing $\nu$
results in a smeared MM$^2$ distribution that is almost triangular
in shape starting near -0.05 GeV$^2$, peaking near 0.10 GeV$^2$, and
ending at 0.75 GeV$^2$.

\begin{figure}[htbp]
\centering \vskip -2mm
\includegraphics[width=2.5in]
{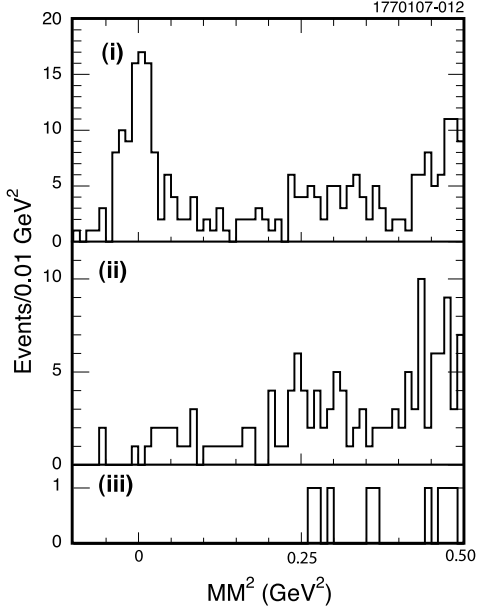} \vskip -3mm
 \caption{The MM$^2$ distributions from data using $D_s^-$ tags, and
 one additional opposite-sign
charged track and no extra energetic showers, for cases (i), (ii),
and (iii). }\label{mm2-data}
 \end{figure}
 The MM$^2$ distributions from data are shown in Fig.~\ref{mm2-data}.
The overall signal region is -0.05 $<$ MM$^2 < 0.20 {\rm ~GeV}^2$. The upper limit is chosen
to prevent background from $\eta\pi^+$ and ${K}^0\pi^+$
final states. The peak in Fig.~\ref{mm2-data}(i) is due to
$D_s^+\to\mu^+\nu$. Below 0.20 GeV$^2$ in both (i) and (ii) we have
$\pi^+\overline{\nu}\nu$ events. The specific signal regions are:
for $\mu^+\nu$, $-0.05<{\rm MM}^2<0.05$ GeV$^2$, corresponding to
$\pm 2\sigma$; for $\pi^+\overline{\nu}\nu$, in case (i) $0.05<{\rm
MM}^2<0.20$ GeV$^2$ and in case (ii) $-0.05<{\rm MM}^2<0.20$
GeV$^2$. In these regions we find 92, 31, and 25 events,
respectively.

We consider backgrounds from two sources: one from real
$D_s^+$ decays and the other from the background under the
single-tag signal peaks. For the latter, we estimate the
background from data using side-bands of the invariant mass, shown
in Fig.~\ref{Inv-mass}.
 For case (i) we find 3.5 (properly normalized) background events in the $\mu^+\nu$
region and 2.5 backgrounds in the $\tau^+\nu$ region; for case
(ii) we find 3 events. Our total background estimate summing over
all of these cases is 9.0$\pm$2.3 events.

The background from real $D_s^+$ decays is evaluated by identifying
specific sources. For $\mu^+\nu$ the only possible background is
$D_s^+\to\pi^+\pi^0$. Using a 195 pb$^{-1}$ subsample of our data,
we limit the branching fraction as $<1.1\times 10^{-3}$ at 90\% C.L.
\cite{PRD}. This low rate coupled with the extra $\gamma$ veto
yields a negligible contribution. The real $D_s^+$ backgrounds for
$\pi^+\overline{\nu}\nu$  are listed in Table~\ref{tab:taunubkrd}.
Using the SM expected ratio of decay rates we calculate a
contribution of 7.4 $\pi^+\overline{\nu}\nu$ events.

\begin{table}[htb]
\begin{center}
\caption{Event backgrounds in the $\pi^+\overline{\nu}\nu$ sample from
real $D_s^+$ decays.
\label{tab:taunubkrd}}
\begin{tabular}{lcccc}
 \hline\hline
    Source  & ${\cal{B}}$(\%)            &   case (i) &   case (ii) & Sum\\ \hline
$D_s^+\to X \mu^+\nu$  & 8.2  & 0$^{+1.8}_{-0}$ & 0 & 0$^{+1.8}_{-0}$\\
$D_s^+\to\pi^+\pi^0\pi^0 $ & 1.0  & 0.03$\pm$0.04 & 0.08$\pm$0.03 & 0.11$\pm$0.04\\
$D_s^+\to\tau^+\nu$ & 6.4 & & &\\
~~~~$\tau^+\to \pi^+\pi^0\overline{\nu}$ & 1.5 & 0.55$\pm$0.22 & 0.64$\pm$0.24 & 1.20$\pm$0.33\\
~~~~$\tau^+\to \mu^+\overline{\nu}\nu$ &1.0& 0.37$\pm$0.15 &  0 & 0.37$\pm$0.15\\
\hline
Sum & & 1.0$^{+1.8}_{-0}$ & 0.7$\pm$0.2 & 1.7$^{+1.8}_{-0.4}$\\
\hline\hline
\end{tabular}
\end{center}
\end{table}

The event yield in the signal region, $N_{\rm det}$ (92), is related
to the number of tags, $N_{\rm tag}$, the branching fractions, and the
background $N_{\rm bkgrd}$ (3.5) as
\begin{eqnarray}
N_{\rm det}-N_{\rm bkgrd}&=&N_{\rm tag}\cdot
\epsilon[\epsilon'{\cal{B}}(D_s^+\to
\mu^+\nu)\label{eq:munuB}\\\nonumber &&+\epsilon''{\cal{B}}(D_s^+\to
\pi^+\overline{\nu}\nu)],\nonumber
\end{eqnarray}
where $\epsilon$ (80.1\%) includes the efficiencies (77.8\%) for
reconstructing the single charged track including final state
radiation, (98.3)\% for not having another unmatched cluster in
the event with energy greater than 300 MeV, and the correction for
the tag bias (4.8\%); $\epsilon'$ (91.4\%) is the product of the
99.0\% $\mu^+$ calorimeter efficiency and the 92.3\% acceptance of
the MM$^2$ cut of $|$MM$^2|< 0.05$ GeV$^2$; $\epsilon''$ (7.6\%)
is the fraction of $\pi^+\overline{\nu}\nu$ events contained in
the $\mu^+\nu$ signal window (13.2\%) times the 60\% acceptance
for a pion to deposit less than 300 MeV in the calorimeter. Using
${\cal{B}}(\tau^+\to\pi^+\overline{\nu}$) of (10.90$\pm$0.07)\%
\cite{PDG}, the ratio of the $\pi^+\overline{\nu}\nu$ to
$\mu^+\nu$ widths is 1.059; we find:
\begin{equation}
\label{eq:bmunurate}
 {\cal{B}}(D_s^+\to \mu^+\nu)= (0.594\pm
0.066\pm0.031)\%.
\end{equation}

We can also sum the $\mu^+\nu$ and $\tau^+\nu$ contributions for
$-0.05<{\rm MM}^2<0.02~{\rm GeV}^2$. Equation~\ref{eq:munuB} still
applies. The number of signal and background events changes to 148
and 10.7, respectively. $\epsilon'$ becomes 96.2\%, and $\epsilon''$
increases to 45.2\%. The effective branching fraction, assuming lepton
universality, is
\begin{equation}
{\cal{B}}^{\rm eff}(D_s^+\to\mu^+\nu) =(0.638 \pm 0.059 \pm 0.033)\%
. \label{eq:finalBR}
\end{equation}

The systematic errors on these branching fractions are dominated by the
error on the number of tags (5\%). Other errors include: (a) track
finding (0.7\%), determined from a detailed comparison of the
simulation with double tag events where one track is ignored; (b)
the error due to the requirement that the charged track deposit no
more than 300 MeV in the calorimeter (1\%), determined using
two-body $D^0\to K^-\pi^+$ decays \cite{DptomunPRD}; (c) the
$\gamma$ veto efficiency (1\%), determined by extrapolating
measurements on fully reconstructed events. Systematic errors
arising from the background estimates are negligible. The total
systematic error for Eq.~\ref{eq:bmunurate} is 5.2\%, and is 5.1\%
for Eq.~\ref{eq:finalBR} as (b) doesn't apply here.

We also analyze the $\tau^+\nu$ final state independently. For case
(i) we define the signal region to be the interval 0.05$<$MM$^2
<$0.20 GeV$^2$, while for case (ii) -0.05$<$MM$^2<$0.20 GeV$^2$.
 The upper limit on MM$^2$ is
 chosen to avoid background from the tail of the ${K}^0\pi^+$ peak.
 The fractions of the MM$^2$
 range accepted are 32\% and 45\% for case (i) and (ii), respectively.

 We find 31 [25] events in the signal region with a background
of 3.5 [5.1] events for case (i) [(ii)]. The branching fraction,
averaging the two cases is
\begin{equation}
{\cal{B}}(D_s^+\to \tau^+\nu)=(8.0\pm 1.3\pm0.4)\% ,
\end{equation}
where the systematic error includes a contribution of 0.06\% from
the uncertainty on ${\cal{B}}(\tau^+\to \pi^+\overline{\nu})$. We
measure $13.4\pm 2.6 \pm 0.2~$  for the ratio of $\tau^+\nu$ to
$\mu^+\nu$ rates using Eq.~\ref{eq:bmunurate}. Here the systematic
error is dominated by the uncertainty on the minimum ionization cut.
We also set an upper limit of ${\cal{B}}(D_s^+\to e^+\nu)< 1.3\times
10^{-4}$ at 90\% C.L. Both of these results are consistent with SM
predictions and lepton universality.

We perform an overall check of our procedures by measuring
${\cal{B}}(D_s^+\to \overline{K}^0K^+)$.  We compute the MM$^2$
(Eq.~\ref{eq:mm2}) using events with an additional charged track
identified as a kaon. These track candidates have momenta of
approximately 1 GeV/$c$; here the RICH has a pion to kaon fake rate
of 1.1\% with a kaon detection efficiency of 88.5\% \cite{fakes}.
For this study, we do not veto events with extra charged tracks, or
$\gamma$'s, because of the presence of the ${K}^0$. We determine
${\cal{B}}(D_s^+\to \overline{K}^0K^+)=(2.90\pm0.19\pm 0.18)\%.$
This method gives a result in good agreement with preliminary CLEO-c
results using double tags of $(3.00\pm 0.19\pm 0.10)$\%
\cite{Peter}; these results are not independent.

We also performed the entire analysis on a MC sample that is 4 times
larger than the data sample. The input branching fraction is 0.5\%
for $\mu^+\nu$ and 6.57\% for $\tau^+\nu$, while our analysis
measured (0.514$\pm$0.027)\% for the case (i) $\mu^+\nu$ signal and
(0.521$\pm$0.024)\% for $\mu^+\nu$ and $\tau^+\nu$ combined.

Using  ${\cal{B}}(D_s^+\to \mu^+\nu)$
 from Eq.~\ref{eq:finalBR},
 and Eq.~\ref{eq:equ_rate} with a $D_s$ lifetime of (500$\pm$7)$\times 10^{-15}\,{\rm s}$ \cite{PDG}, we extract
 \begin{equation}
 f_{D^+_s}=274\pm 13 \pm 7 {~\rm MeV}.
 \end{equation}

We combine with our previous result $f_{D^+}=222.6\pm
16.7^{+2.8}_{-3.4}$ MeV \cite{our-fDp}, and find
\begin{equation}
{f_{D_s^+}}/{f_{D^+}}=1.23\pm 0.11\pm 0.04.
\end{equation}

Lattice QCD predictions for $f_{D_s^+}$ and the ratio
${f_{D_s^+}}/{f_{D^+}}$ have been summarized by Onogi \cite{Onogi}.
Our measurements are consistent with most calculations; examples are
unquenched Lattice that predicts $249 \pm 3 \pm 16 $ MeV and
$1.24\pm 0.01\pm 0.07$ for the ratio \cite{Lat:Milc}, while a recent
quenched prediction gives $266\pm 10 \pm 18$ MeV and $1.13\pm
0.03\pm 0.05$ \cite{Lat:Taiwan}. There is no evidence yet for any
suppression in the ratio due to the presence of a virtual charged
Higgs \cite{Akeroyd}.

The CLEO-c determination of $f_{D_s^+}$ is the most accurate to date
and consistent with other measurements \cite{PDG,PRD}. It also does
not rely on the independent determination of any normalization mode
(e.g. $\phi\pi^+$). (We note that a preliminary CLEO-c result using
$D_s^+\to\tau^+\nu$, $\tau^+\to e^+\overline{\nu}\nu$ \cite{Moscow}
is consistent with these results.)

We gratefully acknowledge the effort of the CESR staff in providing
us with excellent luminosity and running conditions. This work was
supported by the A.P.~Sloan Foundation, the National Science
Foundation, the U.S. Department of Energy, and the Natural Sciences
and Engineering Research Council of Canada.


\begin{thebibliography}{99}

\bibitem{formula-mix}
G. Buchalla, A. J. Buras and M. E. Lautenbacher, Rev. Mod. Phys.
{\bf 68}, 1125 (1996).

\bibitem{CDF}
 A. Abulencia \etal~(CDF), Phys. Rev. Lett. {\bf 97}, 242003 (2006).
See also V. Abazov \etal~(D0), Phys. Rev. Lett. {\bf 97}, 021802
(2006).

\bibitem{Davies}
C. Davies \etal, Phys. Rev. Lett. {\bf 92}, 022001 (2004).

\bibitem{our-fDp}
M. Artuso \etal ~(CLEO), Phys. Rev. Lett. {\bf 95}, 251801 (2005).

\bibitem{DptomunPRD}
G. Bonvicini \etal ~(CLEO) Phys. Rev. {\bf D70}, 112004 (2004).

\bibitem{Formula1}
D. Silverman and H. Yao, Phys. Rev.  {\bf D38}, 214 (1988).

\bibitem{PDG}
W.-M. Yao \etal, J. Phys. {\bf G33}, 1 (2006).

\bibitem{PRD}
T. K. Pedlar \etal~(CLEO), arXiv:0704.0437[hep-ex], submitted to
Phys. Rev. {\bf D}.

\bibitem{Hewett}
J. Hewett, [hep-ph/9505246]; W.-S. Hou, {Phys. Rev.} {\bf D48},
2342 (1993).

\bibitem{Akeroyd}
A. G. Akeroyd, Prog. Theor. Phys. {\bf 111}, 295 (2004).

\bibitem{CLEODR}
D. Peterson \etal , Nucl. Instrum. and Meth. {\bf A478}, 142 (2002);
Y. Kubota \etal , Nucl. Instrum. and Meth. {\bf A320}, 66 (1992).

\bibitem{fakes}
M. Artuso \etal, Nucl. Instrum. Meth. {\bf A554}, 147 (2005).

\bibitem{poling}
R. Poling, [hep-ex/0606016].

\bibitem{taunu}
 P. Rubin \etal ~(CLEO), Phys. Rev. {\bf D73}, 112005 (2006).

\bibitem{Peter}
N.~E.~Adam \etal~(CLEO),  [hep-ex/0607079].

\bibitem{Onogi}
T. Onogi [hep-lat/0610115].

\bibitem{Lat:Milc}
 C. Aubin \etal , Phys. Rev. Lett. {\bf 95}, 122002
 (2005).

 \bibitem{Lat:Taiwan}
 T. W. Chiu \etal, Phys. Lett. {\bf B624}, 31
 (2005).


\bibitem{Moscow}
S. Stone [hep-ex/0610026].
\end{thebibliography}
\end{document}